# How (In)accurate Are Demand Forecasts in Public Works Projects? The Case of Transportation.

**DRAFT 9.0: PLEASE DO NOT QUOTE, COPY, OR CIRCULATE, EXCEPT TO REFEREES**

Bent Flyvbjerg, Mette K. Skamris Holm, and Søren L. Buhl, Aalborg University, Denmark

**All communication to:**

Professor Dr. Bent Flyvbjerg

Department of Development and Planning, Aalborg University

Fibigerstraede 11, 9220 Aalborg, Denmark

Tel: +45 9816 9084 or +45 9635 8080/8379

Fax: +45 9815 3537, E-mail: flyvbjerg@plan.aau.dk

## Biographical Notes

**Bent Flyvbjerg** is a professor of planning at Aalborg University, Denmark. He is founder and director of the university's research program on large-scale transportation infrastructure planning. His latest books are *Megaprojects and Risk* (Cambridge University Press, 2003; with Nils Bruzelius and Werner Rothengatter), *Making Social Science Matter* (Cambridge University Press, 2001), and *Rationality and Power* (University of Chicago Press, 1998). **Mette K. Skamris Holm** is a former assistant professor of planning at Aalborg University. She now works as a planner with Aalborg Municipality. **Søren L. Buhl** is associate professor of mathematics at Aalborg University. He is associate statistician with the university's research program on large-scale transportation infrastructure planning.

## Abstract

This article presents results from the first statistically significant study of traffic forecasts in transportation infrastructure projects. The sample used is the largest of its kind, covering 210 projects in 14 nations worth US$59 billion. The study shows with very high statistical significance that forecasters generally do a poor job of estimating the demand for transportation infrastructure projects. The result is substantial downside financial and economic risks. Such risks are typically ignored or downplayed by planners and decision makers, to the detriment of social and economic welfare. For nine out of ten rail projects passenger forecasts are overestimated; average overestimation is 106 percent. This results in large benefit shortfalls for rail projects. For half of all road projects the difference between actual and forecasted traffic is more than ±20 percent. Forecasts have not become more accurate over the 30-year period studied. If techniques and skills for arriving at accurate demand forecasts have improved over time, as often claimed by forecasters, this does not show in the data. The causes of inaccuracy in forecasts are different for rail and road projects, with political causes playing a larger role for rail than for road. The cure is transparency, accountability, and new forecasting methods. The challenge is to change the governance structures for forecasting and project development. The article shows how planners may help achieve this.

## Introduction

Despite the enormous sums of money being spent on transportation infrastructure, surprisingly little systematic knowledge exists about the costs, benefits, and risks involved. The literature lacks statistically valid answers to the central and self-evident question of whether transportation infrastructure projects perform as forecasted. When a project underperforms, this is often explained away as an isolated instance of unfortunate circumstance; it is typically not seen as the particular expression of a general pattern of underperformance in transportation infrastructure projects. Because knowledge is wanting in this area of research, until now it has been impossible to validly refute or confirm whether underperformance is the exception or the rule.

In three previous articles we answered the question of project performance as regards costs and cost-related risks. We found that projects do not perform as forecasted in terms of costs; almost nine out of ten projects fall victim to significant cost overrun. We also investigated the causes and cures of such underperformance (Flyvbjerg, Holm, and Buhl 2002, 2003, 2004; see also Flyvbjerg, Bruzelius, and Rothengatter 2003). In this article we focus on the benefit-side of investments and answer the question of whether projects perform as forecasted in terms of demand and revenue risks. We compare forecasted performance in terms of demand with actual performance for a large number of projects. Knowledge about cost risk, benefit risk, and compound risk is crucial to making informed decisions about projects. This is not to say that costs and benefits are or should be the only basis for deciding whether to build or not. Clearly other forms of rationality than economic rationality are at work in most infrastructure projects and are balanced in the broader frame of public decision making. But the costs and benefits of infrastructure projects often run in the hundreds of millions of dollars, with risks being correspondingly high. Without knowledge of such risks, decisions are likely to be flawed.

As pointed out by Pickrell (1990) and Richmond (1998), estimates of the financial viability of projects are heavily dependent on the accuracy of traffic demand forecasts. Such forecasts are also the basis for socio-economic and environmental appraisal of transportation infrastructure projects. According to the experiences gained with the accuracy of demand forecasting in the transportation sector, covering traffic volumes, spatial traffic distribution, and distribution between transportation modes, there is evidence that demand forecasting--like cost forecasting, and despite all scientific progress in modeling--is a major source of uncertainty and risk in the appraisal of transportation infrastructure projects.

Traffic forecasts are routinely used to dimension the construction of transportation infrastructure projects. Here accuracy in forecasts is a point of considerable importance to the effective allocation of scarce funds. For example, Bangkok's US$2 billion Skytrain was hugely overdimensioned because the passenger forecast were 2.5 times higher than actual

traffic. As a result, station platforms are too long for the shortened trains that now operate the system, a large number of trains and cars are idly parked in the train garage because there is no need for them, terminals are too large, etc. The project company has ended up in financial trouble and even though urban rail is probably a good idea for a congested and air-polluted city like Bangkok, overinvesting in idle capacity is hardly the best way to use resources, and especially not in a developing nation where capital for investment is scarce. Conversely, a UK National Audit Office study identified a number of road projects that were underdimensioned because traffic forecasts were too low. This, too, led to multi-million-dollar inefficiencies, because it is much more expensive to add capacity to existing fully used roads than it is to build the capacity up front (National Audit Office 1988). For these and other reasons, accuracy in traffic forecasts matter.

Nevertheless, rigorous studies of accuracy are rare. Where such studies exist, they are characteristically small-N research, that is, they are single-case studies or they cover only a sample of projects too small or too uneven to allow systematic, statistical analyses (Brooks and Trevelyan 1979, Fouracre et al. 1990, Fullerton and Openshaw 1985, Kain 1990, Mackinder and Evans 1981, National Audit Office 1988 and 1992, Pickrell 1990, Richmond 1998, Walmsley and Pickett 1992, Webber 1976, World Bank 1994). Despite their value in other respects, with these and other studies, it has so far been impossible to give statistically satisfying answers to questions about how accurate traffic forecasts are for transportation infrastructure projects.

The objective of the present study has been to change this state of affairs by establishing a sample of transportation infrastructure projects that is sufficiently large to permit statistically valid answers to questions of accuracy. In addition to this objective, it has been a practical objective to give planners the tools for carrying out realistic and valid risk assessment of projects as regards travel demand. Existing studies almost all conclude there is a strong tendency for traffic forecasts to be overestimated (Mackinder and Evans 1981: 25; National Audit Office 1985: app. 5.16; World Bank 1986; Fouracre et al. 1990: 1, 10; Pickrell 1990: x; Walmsley and Pickett 1992: 2; Thompson 1993: 3-4). Below we will show that this conclusion is a consequence of the small samples used in existing studies; it does not hold for the project population. When we enlarge the sample of projects by a factor 10-20 to a more representative one, we find a different picture. Road traffic forecasts are not generally overestimated, although they are often very inaccurate, whereas forecasts of rail patronage are generally overestimated, to be sure.

We follow common practice and define the inaccuracy of a traffic forecast as actual minus forecasted traffic in percentage of forecasted traffic. Actual traffic is counted for the first year of operations (or the opening year). Forecasted traffic is the traffic estimate for the first year of operations (or the opening year) as estimated at the time of decision to build the project. Thus the forecast is the estimate available to decision makers when they made the decision to build the project in question. If no estimate was available at the time of decision to build, then the closest available estimate was used, typically a later estimate resulting in a conservative bias in our measure for inaccuracy.

Inaccuracy of traffic forecasts is measured for each project in a sample of 210 transportation infrastructure projects with comparable data for forecasted and actual traffic. The sample comprises a project portfolio worth approximately US$59 billion in actual costs (2004 prices). The portfolio includes 27 rail projects and 183 road projects completed between 1969 and 1998. The project types are urban rail, high-speed rail, conventional rail, bridges, tunnels, highways, and freeways. The projects are located in 14 countries on 5 continents, including both developed and developing nations. The 14 countries are: Brazil, Chile, Denmark, Egypt, France, Germany, Hong Kong, India, Mexico, South Korea, Sweden, Tunisia, UK, USA. Projects were selected for the sample on the basis of data availability and quality.[1] As far as we know, this is the largest sample of projects with comparable data on forecasted and actual traffic that has been established for this type of project. For a full description of the sample, data, and methods of testing for inaccuracy, please see Flyvbjerg (2004).

## Are Rail or Road Forecasts More Accurate?

Figures 1 and 2 show the distribution of inaccuracy of traffic forecasts for the 210 projects in the sample split into rail and road projects. Perfect accuracy is indicated by zero; a negative figure indicates that actual traffic is that many percent lower than forecasted traffic; a positive figure indicates that actual traffic is that many percent higher than forecasted traffic. The most noticeable attribute of Figures 1 and 2 is the striking difference between rail and road projects. Rail passenger forecasts are much more inaccurate (inflated) than are road traffic forecasts.

[Figures 1-2 app. here]

Tests show that of the 27 rail projects included in the statistical analyses, two German projects should be considered as statistical outliers. These are the two projects represented by the two rightmost columns in the rail histogram in Figure 1 and the two uppermost plots in the rail box-plot diagram shown in Figure 2. Excluding statistical outliers, we find the following results for the remaining 25 rail projects (results including the two statistical outliers are given in square parentheses):

• The data document a massive problem with inflated rail passenger forecasts. For more than 9 out of 10 rail projects passenger forecasts are overestimated; for 72 percent of all rail projects, passenger forecasts are overestimated by more than two thirds. [Including statistical outliers: For 67 percent of all rail projects, passenger forecasts are overestimated by more than two thirds].

• Rail passenger forecasts were overestimated by an average of 105.6 percent (95 percent confidence interval of 66.0 to 169.9), resulting in actual traffic that was on average 51.4 percent lower than forecasted traffic (sd=28.1, 95 percent confidence interval of -62.9 to -39.8). [Including statistical outliers: Rail passenger forecasts were overestimated by an average of 65.2 percent (95 percent confidence interval of 23.1 to 151.3), resulting in actual traffic that was on average 39.5 percent lower than forecasted traffic (sd=52.4, 95 percent confidence interval of -60.2 to -18.8)].

- 84 percent of the rail projects have actual traffic more than 20 percent below forecasted traffic and none have actual traffic more than 20 percent above forecasted traffic. Even if we double the threshold value to 40 percent, we find that a solid 72 percent of all rail projects have actual traffic below that limit. [Including statistical outliers the figures are 85 percent and 74 percent, respectively.]

[Table 1 app. here]

For road projects, we find with 95 percent confidence that there is no significant difference (p=0.638) in terms of forecast inaccuracies between vehicle traffic on highways, bridges, and in tunnels (170 highways, 10 bridges, 3 tunnels). Hence we consider the 183 road projects as an aggregate. Our tests show (see also Table 1):

- 50 percent of the road projects have a difference between actual and forecasted traffic of more than ±20 percent. If we double the threshold value to ±40 percent, we find that 25 percent of projects are above this level.

- There is no significant difference between the frequency of inflated versus deflated forecasts for road vehicle traffic (p=0.822, two-sided binominal test). 21.3 percent of projects have inaccuracies below -20 percent, whereas 28.4 percent of projects have inaccuracies above +20 percent.

- Road traffic forecasts were underestimated by an average of 8.7 percent (95 percent confidence interval of 2.9 to 13.7), resulting in actual traffic that was on average 9.5 percent higher than forecasted traffic (sd=44.3, 95 percent confidence interval of 3.0 to 15.9).

Here it would be interesting to compare toll roads with non-toll roads, but unfortunately the present data do not allow this.

We see that the risk is substantial that road traffic forecasts are wrong by a large margin, but the risk is more balanced than for rail passenger forecasts. Testing the difference between rail and road, we find at a very high level of statistical significance that rail passenger forecasts are less accurate and more inflated than road vehicle forecasts ($p < 0.001$, Welch two-sample t-test). However, there is no indication of a significant difference between the standard deviations for rail and road forecasts, both are high, indicating a large element of uncertainty and risk for both types of forecasts ($p = 0.213$, two-sided F-test). Excluding the two statistical outliers for rail, we find the standard deviation for rail projects to be significantly lower than for road projects, although still high ($p = 0.0105$).

Any traffic forecast is done in the context of uncertainty about many of the key inputs and drivers of the projection--demographics, economic factors, technology, and differences between the assumed and actual operating service plans that are implemented. The same holds for other important aspects of project evaluation and investment decision making, including forecasts of costs (Flyvbjerg, Holm, and Buhl 2002, 2003, 2004). Simple uncertainty would account for the type of inaccuracy we find with road traffic forecasts, with a fairly even distribution of high and low forecasts. Simple uncertainty does not seem to account for the outcome of rail travel forecasts, however. Such forecasts are overestimated too consistently for an interpretation in terms of simple uncertainty to be statistically plausible.

We conclude that the traffic estimates used in decision making for rail infrastructure development are highly, systematically, and significantly misleading (inflated). The result is large benefit shortfalls. For road projects the problem of misleading forecasts is less severe and less one-sided than for rail. But even for roads, for half the projects the difference between actual and forecasted traffic is more than ±20 percent. On this background, planners and decision makers are well advised to take with a grain of salt any traffic forecast which does not explicitly take into account the uncertainty of predicting future traffic. For rail passenger forecasts, a grain of salt may not be enough. The data demonstrate to planners that risk assessment and management regarding travel demand must be an integral part of planning for both rail and road projects. This is especially the case because prediction errors in the early stages of forecasting appear to amplify, rather than decrease, in later stages (Zhao and Kockelman 2001, Mierzejewski 1995). The data presented above provide the empirical basis on which planners may establish risk assessment and management. Below we propose methods and procedures for such risk assessment and management.

## Have Forecasts Become More Accurate Over Time?

Figures 3 and 4 show how forecast inaccuracy varies over time for the projects in the sample for which inaccuracy could be coupled with information about year of decision to build and/or year of completing the project. Statistical tests show there is no indication that traffic forecasts have become more accurate over time, despite claims to the opposite (American Public Transit Association 1990, 6, 8). For road projects, forecasts even appear to become more inaccurate toward the end of the 30-year period studied. Statistical analyses corroborate this impression.

[Figures 3-4 app. here]

For rail projects, forecast inaccuracy is independent of both year of project commencement or year of project conclusion. This is the case whether the two German projects (marked with "K" in Figure 3) are treated as statistical outliers or not. We conclude that forecasts of rail passenger traffic have not improved over time. Rail passenger traffic has been consistently overestimated during the 30-year period studied. The US Federal Transit Administration, FTA, has a study underway which indicates that rail passenger forecasts may have become more accurate recently (Ryan 2004). According to an oral presentation of the study at the Annual TRB Meeting in 2004, of 19 new rail projects 68 percent achieved actual patronage less than 80 percent of forecast patronage. This is a 16 percentage points improvement over the rail projects in our sample, where 84 percent of rail projects achieved actual patronage less than 80 percent of that forecasted (see above). It is also an improvement over the situation Pickrell (1990) depicted.[2] It is unclear, however, whether this reported improvement is statistically significant, and despite the improvement the same pattern of overestimation continues. Ryan's (2004, slide 30) preliminary conclusion thus dovetails with ours: "Risk of large errors still remains." A report from the FTA study is expected at the end of 2004.

For road projects, inaccuracies are larger towards the end of the period with highly underestimated traffic. However, there is a difference between Danish and other road projects. For Danish road projects, we find at a very high level of statistical significance that inaccuracy varies with time ($p<0.001$). After 1980 Danish road traffic forecasts offered large underestimations, whereas this was not the case for Denmark before 1980, nor was it the case for other countries for which data exist. During a decade from the second half of the 1970's to the second half of the 1980's, inaccuracy of Danish road traffic forecasts increased 18-fold, from 3 to 55 percent (see Figure 5).

[Figure 5 app. here]

The Danish experience with increasing inaccuracy in road traffic forecasts is best explained by what Ascher (1979: 52, 202-203) calls "assumption drag," that is, the continued use of assumptions after their validity has been contradicted by the data. More specifically, traffic forecasters typically calibrate forecasting models on the basis of data from the past. The so-called energy crises of 1973 and 1979 and associated increases in petrol prices plus decreases in real wages had a profound, if short-lived, effect on road traffic in Denmark, with traffic declining for the first time in decades. Danish traffic forecasters adjusted and calibrated their models accordingly on the assumption that they were witnessing an enduring trend. The assumption was mistaken. When, during the 1980s, the effects of the two oil crises and related policy measures tapered off, traffic boomed again rendering forecasts made on 1970's assumptions inaccurate.

We conclude that accuracy in traffic forecasting has not improved over time. Rail passenger forecasts are as inaccurate, that is, inflated, today as they were 30 years ago. Road vehicle forecasts even appear to have become more inaccurate over time with large underestimations towards the end of the 30-year period studied. If techniques and skills for arriving at accurate traffic forecasts have improved over time, this does not show in the data. This suggests to planners that the most effective means for improving forecasting accuracy is probably not improved models but, instead, more realistic assumptions and systematic use of empirically based assessment of uncertainty and risk. Below, in the section on reference class forecasting, we will see how this may be done. For rail, in particular, the persistent existence over time of highly inflated passenger forecasts invites speculation that an equilibrium has been reached where strong incentives and weak disincentives for overestimating passenger traffic may have taught project promoters that overestimated passenger forecasts pay off: in combination with underestimated costs such forecasts help misrepresent rail projects to decision makers in ways that help get rail projects approved and built (Flyvbjerg, Bruzelius, and Rothengatter 2003). This suggests that improved accuracy for rail forecasts will require strong measures of transparency and accountability to curb strategic misrepresentation in forecasts. Such measures form part of what has become known as PPPs - public-private partnerships - and there is some indication that properly designed PPPs may help improve the accuracy of cost forecasts (National Audit Office 2003). As far as we know, no studies exist regarding the effect of PPPs or similar arrangements on the accuracy of traffic forecasts.

## Does Project Size, Length of Implementation, and Geography Matter to Accuracy?

Testing for effect on forecasting inaccuracy from size of project, we used linear regression analyses measuring size of project by estimated costs, estimated number of passengers, and estimated number of vehicles.[3] As the distributions of estimated costs, estimated number of passengers, and estimated number of vehicles are all skew, the logarithms of these have also been used as explanatory variables.

For rail projects, based on 17 cases we found that inaccuracies in passenger forecasts are not significantly dependent on costs (p=0.177), but do have significance dependent on logarithm of costs (p=0.018), with higher costs leading to higher inaccuracies. Based on 27 cases, inaccuracies in passenger forecasts are not significantly dependent on estimated size of number of passengers, neither directly (p=0.738) nor taking logarithms (p=0.707).

For road projects, based on 24 cases, inaccuracies in vehicle forecast are not significantly dependent on costs, neither directly (p=0.797) nor logarithmically (p=0.114). Based on 51 cases, inaccuracies in vehicle forecast are significantly dependent on estimated number of vehicles, both directly (p=0.011) and even stronger taking logarithms (p<0.001), with smaller projects tending to have the most inaccurate, underestimated, traffic forecasts.

We know of only one other study that relates inaccuracy in travel demand forecasting with size of project (Maldonado 1990, quoted in Mierzejewski 1995: 31). Based on data from 22 US airports, this study found that inaccuracy in aviation forecasting did not correlate with size of facility.

Additional tests indicate no effect on inaccuracy from length of project implementation phase, defined as the time period from decision to build a project until operations begin. More data are needed in order to study the effect on inaccuracy from geographic location of projects and type of ownership. With the available data, there is no significant difference between geographical areas, which suggests that until such a time when more data are available, planners may pool data from different geographical areas when carrying out risk assessment.

## Causes of Inaccuracies and Bias in Traffic Forecasts

The striking difference in forecasting inaccuracy between rail and road projects documented above may possibly be explained by the different procedures that apply to how each type of project is funded, where competition for funds are typically more pronounced for rail than for road, which creates an incentive for rail promoters to present their project in as favorable a light as possible, that is, with overestimated benefits and underestimated costs (see more in Flyvbjerg, Holm, and Buhl 2002). One may further speculate that rail patronage will be overestimated and road traffic underestimated in instances where there is a strong political or ideological desire to see passengers shifted from road to rail, for instance for reasons of congestion or protection of the environment. Forecasts here become part of the political rhetoric aimed at showing voters that something is being done--or will be done--about the problems at hand. In such cases it may be difficult for forecasters and planners to argue for more realistic forecasts, because politicians here use forecasts to show political intent, not the most likely outcome.

In order to arrive at a more systematic analysis of causes of inaccuracies in traffic forecasts, we identified such causes for 234 transportation infrastructure projects. For a number of projects we were able to identify causes of inaccuracies but not the numerical size of inaccuracies. This explains why we have more projects (234) in this part of our analysis than in the previous part (210 projects).[4] Causes of inaccuracies are stated causes that explain differences between actual and forecasted traffic for the first year of operations or the opening year. For the projects for which we did the data collection, project managers were asked to account for the factors that would explain why actual traffic was different from forecasted traffic. For the other projects the stated causes are a mixture of this type of statement by managers supplemented by statements by researchers about what caused such differences. For these projects, the data do not allow an exact distinction between manager statements and researcher statements, even though such a distinction would be desirable. It is a problem with using stated causes that what people say they do is often significantly different from what they actually do. Uncovering revealed causes for inaccuracy in traffic forecasting is therefore an important area for further research. For the time being we have to make do with stated causes.

Figure 6 shows the stated causes for inaccuracies in traffic forecasts for rail and road, respectively. For each transportation mode and stated cause, a column shows the percentage of projects for which this cause was stated as a reason for inaccuracy.

[Figure 6 app. here]

Again the results are very different for rail and road. For rail projects, the two most important stated causes are "uncertainty about trip distribution" and "deliberately slanted forecasts." Trip distribution in rail passenger models, while ideally based on cross-sectional data collected from users of transportation system, is often adapted to fit national or urban policies aimed at boosting rail traffic. Here too, it is difficult for forecasters and planners to gain acceptance for realistic forecasts that run counter to idealistic policies. But such policies frequently fail and the result is the type of overestimated passenger forecast which we have documented above as typical for rail passenger forecasting (Flyvbjerg, Bruzelius, and Rothengatter 2003, ch. 3). As regards deliberately slanted forecasts, such forecasts are produced by rail promoters in order to increase the likelihood that rail projects get built (Wachs 1990). Such forecasts exaggerate passenger traffic and thus revenues. Elsewhere we have shown that the large overestimation of traffic and revenues documented above for rail goes hand-in-hand with an equally large underestimation of costs (Flyvbjerg, Holm, and Buhl 2002, 2004). The result is cost-benefit analyses of rail projects that are inflated, with benefit-cost ratios that are useful for getting projects accepted and built.

For road projects, the two most often stated causes for inaccurate traffic forecasts are uncertainties about "trip generation" and "land-use development." Trip generation is based on traffic counts and demographic and geographical data. Such data are often dated and incomplete and forecasters quote this as a main source of uncertainty in road traffic forecasting. Forecasts of land-use development are based on land-use plans. The land-use actually implemented is often quite different from what was planned, however. This, again, is a source of uncertainty in forecasting.

The different patterns in stated causes for rail and road, respectively, fit well with the figures for actual forecast inaccuracies documented above. Rail forecasts are systematically and significantly overestimated to a degree that indicates intent and not error on the part of rail forecasters and promoters. The stated causes, with "deliberately slanted forecasts" as the second to largest category, corroborate this interpretation, which corresponds with findings by Wachs (1986); Flyvbjerg, Holm, and Buhl (2002); and the UK Department for Transport (2004, 43-55). Road forecasts are also often inaccurate, but they are substantially more balanced than rail forecasts, which indicate a higher degree of fair play in road forecasting. This interpretation is corroborated by the fact that deliberately slanted forecasts are not quoted as a main cause of inaccuracy for road traffic forecasts, whereas more technical factors like trip generation and land-use development are. This is not to say that road traffic forecasts are never politically manipulated. It is to say, however, that this appears to happen less often and less systematically for road than for rail projects. It is also not to say that road projects generally have a stronger justification than rail projects; just that they have less biased forecasts than rail projects.

## What Planners Can Do to Reduce Inaccuracy, Bias, and Risk in Forecasting

The results presented above show that it is highly risky to rely on travel demand forecasts to plan and implement large transportation infrastructure investments. Rail passenger forecasts are overestimated in 9 out of 10 cases, with an average overestimate above 100 percent. Half of all road traffic forecasts are wrong by more than ±20 percent. Forecasts have not become more accurate for 30 years. This state of affairs points directly to better risk assessment and management as something planners could and should do to improve planning and decision making for transportation infrastructure projects. Today, the benefit risks generated by inaccurate travel demand forecasts are widely ignored or underestimated in planning, just as cost risks are neglected (Flyvbjerg, Holm, and Buhl 2003).

When contemplating what planners can do to reduce inaccuracy, bias, and risk in forecasting, we need to distinguish between two fundamentally different situations:

Situation 1: Planners consider it important to get forecasts right.

Situation 2: Planners do not consider it important to get forecasts right, because optimistic forecasts are seen as a means to getting projects started.

We consider the first situation in this section and the second in the following section.

If planners genuinely consider it important to get forecasts right, we recommend they use a new forecasting method called "reference class forecasting" to reduce inaccuracy and bias. This method was originally developed to compensate for the type of cognitive bias in human forecasting that Princeton psychologist Daniel Kahneman found in his Nobel prize-winning work on bias in economic forecasting (Kahneman 1994, Kahneman and Tversky 1979). Reference class forecasting has proven more accurate than conventional forecasting. For reasons of space, here we present only an outline of the method, based mainly on Lovallo and Kahneman (2003) and Flyvbjerg (2003). In a different context, we are currently developing what is, to our knowledge, the first instance of practical reference class forecasting in planning (UK Department for Transport 2004).

Reference class forecasting consists in taking a so-called "outside view" on the particular project being forecast. The outside view is established on the basis of information from a class of similar projects. The outside view does not try to forecast the specific uncertain events that will affect the particular project, but instead places the project in a statistical distribution of outcomes from this class of reference projects. Reference class forecasting requires the following three steps for the individual project:

(1)     Identification of a relevant reference class of past projects. The class must be broad enough to be statistically meaningful but narrow enough to be truly comparable with the specific project.

(2)     Establishing a probability distribution for the selected reference class. This requires access to credible, empirical data for a sufficient number of projects within the reference class to make statistically meaningful conclusions.

(3)     Compare the specific project with the reference class distribution, in order to establish the most likely outcome for the specific project.

Daniel Kahneman relates the following story about curriculum planning to illustrate reference class forecasting in practice (Lovallo and Kahneman 2003, 61). We use this example, because similar examples do not exist as yet in the field of transportation planning. Some years ago, Kahneman was involved in a project to develop a curriculum for a new subject area for high schools in Israel. The project was carried out by a team of academics and teachers. In time, the team began to discuss how long the project would take to complete. Everyone on the team was asked to write on a slip of paper the number of months needed to finish and report the project. The estimates ranged from 18 to 30 months. One of the team members--a distinguished expert in curriculum development--was then posed a challenge by another team member to recall as many projects similar to theirs as possible and to think of these projects as they were in a stage comparable to their project. "How long did it take them at that point to reach completion?", the expert was asked. After a while he answered, with some discomfort, that not all the comparable teams he could think of ever did complete their task. About 40 percent of them eventually gave up. Of those remaining, the expert could not think of any that completed their task in less than seven years, nor of any that took more than ten. The expert was then asked if he had reason to believe that the present team was more skilled in curriculum development than the earlier ones had been. The expert said no, he did not see any relevant factor that distinguished this team favorably from the teams he had been thinking about. His impression was that the present team was slightly below average in terms of resources and potential. The wise decision at this point would probably have been for the team to break up, according to Kahneman. Instead, the members ignored the pessimistic information and proceeded with the project. They finally completed the project eight years later, and their efforts went largely wasted--the resulting curriculum was rarely used.

In this example, the curriculum expert made two forecasts for the same problem and arrived at very different answers. The first forecast was the inside view; the second was the outside view, or the reference class forecast. The inside view is the one that the expert and the other team members adopted. They made forecasts by focusing tightly on the case at hand, considering its objective, the resources they brought to it, and the obstacles to its completion. They constructed in their minds scenarios of their coming progress and extrapolated current trends into the future. The resulting forecasts, even the most conservative ones, were overly optimistic. The outside view is the one provoked by the question to the curriculum expert. It completely ignored the details of the project at hand, and it involved no attempt at forecasting the events that would influence the project's future course. Instead, it examined the experiences of a class of similar projects, laid out a rough distribution of outcomes for this reference class, and then positioned the current project in that distribution. The resulting forecast, as it turned out, was much more accurate.

Similarly--to take an example from city planning--planners in a city preparing to build a new subway would, first, establish a reference class of comparable projects. This could be the urban rail projects included in the sample for this article. Through analyses the planners would establish that the projects included in the reference class were indeed comparable. Second, if the planners were concerned about getting patronage forecasts right, they would then establish the distribution of outcomes for the reference class regarding the accuracy of patronage forecasts. This distribution would look something like the rail part of Figure 1. Third, the planners would compare their subway project to the reference class distribution. This would make it clear to the planners that unless they had reason to believe they are substantially better forecasters and planners than their colleagues who did the forecasts and planning for projects in the reference class, they are likely to grossly overestimate patronage. Finally, planners may then use this knowledge to adjust their forecasts for more realism.

The contrast between inside and outside views has been confirmed by systematic research (Gilovich, Griffin, and Kahneman 2002). The research shows that when people are asked simple questions requiring them to take an outside view, their forecasts become significantly more accurate. However, most individuals and organizations are inclined to adopt the inside view in planning major initiatives. This is the conventional and intuitive approach. The traditional way to think about a complex project is to focus on the project itself and its details, to bring to bear what one knows about it, paying special attention to its unique or unusual features, trying to predict the events that will influence its future. The thought of going out and gathering simple statistics about related cases seldom enters a planner's mind. This is the case in general, according to Lovallo and Kahneman (2003, 61-62). And it is certainly the case for travel demand forecasting. Despite the many forecasts we have reviewed, for instance for this article, we have not come across a single genuine reference class forecast of travel demand.[5] If our readers have information about such forecasts, we would appreciate their feedback for our on-going work on this issue.

While understandable, planners' preference for the inside view over the outside view is unfortunate. When both forecasting methods are applied with equal skill, the outside view is much more likely to produce a realistic estimate. That is because it bypasses cognitive and organizational biases such as appraisal optimism and strategic misrepresentation and cuts directly to outcomes. In the outside view planners and forecasters are not required to make scenarios, imagine events, or gauge their own and others' levels of ability and control, so they cannot get all these things wrong. Surely the outside view, being based on historical precedent, may fail to predict extreme outcomes, that is, those that lie outside all historical precedents. But for most projects, the outside view will produce more accurate results. In contrast, a focus on inside details is the road to inaccuracy.

The comparative advantage of the outside view is most pronounced for non-routine projects, understood as projects that planners and decision makers in a certain locale have never attempted before--like building an urban rail system in a city for the first time, or a new major bridge or tunnel where none existed before. It is in the planning of such new efforts that the biases toward optimism and strategic misrepresentation are likely to be largest. To be sure, choosing the right reference class of comparative past projects becomes more difficult when planners are forecasting initiatives for which precedents are not easily found, for instance the introduction of new and unfamiliar technologies. However, most large-scale transportation projects are both non-routine locally and use well-known technologies. Such projects are, therefore, particularly likely to benefit from the outside view and reference class forecasting. The same holds for concert halls, museums, stadiums, exhibition centers, and other local one-off projects.

## When Planners Are Part of the Problem, Not the Solution

In the present section we consider the situation where planners and other influential actors do not find it important to get forecasts right and where planners, therefore, do not help to clarify and mitigate risk but, instead, generate and exacerbate it. Here planners are part of the problem, not the solution. This situation may need some explication, because it possibly sounds to many like an unlikely state of affairs. After all, it may be agreed that planners ought to be interested in being accurate and unbiased in forecasting. It is even stated as an explicit requirement in the AICP Code of Ethics and Professional Conduct that "A planner must strive to provide full, clear and accurate information on planning issues to citizens and governmental decision-makers" (American Planning Association 1991, A.3), and we certainly agree with the Code. The British RTPI has laid down similar obligations for its members (Royal Town Planning Institute 2001).

However, the literature is replete with things planners and planning "must" strive to do, but which they don't. Planning must be open and communicative, but often it is closed. Planning must be participatory and democratic, but often it is an instrument to dominate and control. Planning must be about rationality, but often it is about power (Flyvbjerg 1998, Watson 2003). This is the "dark side" of planning and planners identified by Flyvbjerg (1996) and Yiftachel (1998), which is remarkably underexplored by planning researchers and theorists.

Forecasting, too, has its dark side. It is here "planners lie with numbers," as Wachs (1989) has aptly put it. Planners on the dark side are busy, not with getting forecasts right and following the AICP Code of Ethics, but with getting projects funded and built. And accurate forecasts are often not an effective means for achieving this objective. Indeed, accurate forecasts may be counterproductive, whereas biased forecasts may be effective in competing for funds and securing the go-ahead for construction. "The most effective planner," says Wachs (1989, 477), "is sometimes the one who can cloak advocacy in the guise of scientific or technical rationality." Such advocacy would stand in direct opposition to AICP's ruling that "the planner's primary obligation [is] to the public interest" (American Planning Association 1991, B.2). Nevertheless, seemingly rational forecasts that underestimate costs and overestimate benefits have long been an established formula for project approval (Flyvbjerg, Bruzelius, and Rothengatter 2003). Forecasting is here mainly another kind of rent-seeking behavior, resulting in a make-believe world of misrepresentation which makes it extremely difficult to decide which projects deserve undertaking and which do not. The consequence is, as even one of the industry's own organs, the Oxford-based Major Projects Association, acknowledges, that too many projects proceed that should not. We would like to add that many projects don't proceed that probably should, had they not lost out to projects with "better" misrepresentation (Flyvbjerg, Holm, and Buhl 2002).

In this situation, the question is not so much what planners can do to reduce inaccuracy and risk in forecasting, but what others can do to impose on planners the checks and balances that would give planners the incentive to stop producing biased forecasts and begin to work according to their Code of Ethics. The challenge is to change the power relations, which governs forecasting and project development. Here better forecasting techniques and appeals to ethics won't do; institutional change with a focus on transparency and accountability is necessary.

Two basic types of accountability define liberal democracies: (1) Public sector accountability through transparency and public control, and (2) Private sector accountability via competition and market control. Both types of accountability may be effective tools to curb planners' misrepresentation in forecasting and to promote a culture which acknowledges and deals effectively with risk. In order to achieve accountability through *transparency and public control*, the following would be required as practices embedded in the relevant institutions:

- National-level government should not offer discretionary grants to local infrastructure agencies for the sole purpose of building a specific type of infrastructure, for instance rail. Such grants create perverse incentives. Instead, national government should simply offer "infrastructure grants" or "transportation grants" to local governments, and let local political officials spend the funds however they choose to, but make sure that every dollar they spend on one type of infrastructure reduces their ability to fund another.

- Forecasts should be made subject to independent peer review. Where large amounts of taxpayers' money are at stake, such review may be carried out by national or state accounting and auditing offices, like the General Accounting Office in the US or the National Audit Office in the UK, who have the independence and expertise to produce such reviews. Other types of independent review bodies may be established, for instance within national departments of finance or with relevant professional bodies.

- Forecasts should be benchmarked against comparable forecasts, for instance using reference class forecasting as described in the previous section.

- Forecasts, peer reviews, and benchmarkings should be made available to the public as they are produced, including all relevant documentation.

- Public hearings, citizen juries, and the like should be organized to allow stakeholders and civil society to voice criticism and support of forecasts. Knowledge generated in this way should be integrated in planning and decision making.

- Scientific and professional conferences should be organized where forecasters would present and defend their forecasts in the face of colleagues' scrutiny and criticism.

- Projects with inflated benefit-cost ratios should be reconsidered and stopped if recalculated costs and benefits do not warrant implementation. Projects with realistic estimates of benefits and costs should be rewarded.

- Professional and occasionally even criminal penalties should be enforced for planners and forecasters who consistently and foreseeably produce deceptive forecasts. An example of a professional penalty would be the exclusion from one's professional organization if one violates its code of ethics. An example of a criminal penalty would be punishment as the result of prosecution before a court or similar legal set-up, for instance where deceptive forecasts have led to substantial mismanagement of public funds (Garett and Wachs, 1996).

Malpractice in planning should be taken as seriously as it is in other professions. Failing to do this amounts to not taking the profession of planning seriously.

In order to achieve accountability in forecasting via *competition and market control*, the following would be required, again as practices that are both embedded in and enforced by the relevant institutions:

- The decision to go ahead with a project should, where at all possible, be made contingent on the willingness of private financiers to participate without a sovereign guarantee for at least one third of the total capital needs.[6] This should be required whether projects pass the market test or not, that is, whether projects are subsidized or not or provided for social justice reasons or not. Private lenders, shareholders, and stock market analysts would produce their own forecasts or would critically monitor existing ones. If they were wrong about the forecasts, they and their organizations would be hurt. The result would be more realistic forecasts and reduced risk.

- Full public financing or full financing with a sovereign guarantee should be avoided.

- Forecasters and their organizations must share financial responsibility for covering benefit shortfalls (and cost overruns) resulting from misrepresentation and bias in forecasting.

- The participation of risk capital should not mean that government gives up or reduces control of the project. On the contrary, it means that government can more effectively play the role it should be playing, namely as the ordinary citizen's guarantor for ensuring concerns about safety, environment, risk, and a proper use of public funds.

If the institutions with responsibility for developing and building major transportation infrastructure project would effectively implement, embed, and enforce such measures of accountability, then the misrepresentation in transportation forecasting, which is widespread today, may be mitigated. If this is not done, misrepresentation is likely to continue, and the allocation of funds for transportation investments is likely to be wasteful.

## Conclusions

We conclude that the patronage estimates used by planners of rail infrastructure development are highly, systematically, and significantly misleading (inflated). This results in large benefit shortfalls for rail projects. For road projects the problem of misleading forecasts is less severe and less one-sided than for rail. But even for roads, for half the projects the difference between actual and forecasted traffic is more than ±20 percent. On this background, planners and decision makers are well advised to take with a grain of salt any traffic forecast which does not explicitly take into account the uncertainty of predicting future traffic. For rail passenger forecasts, a grain of salt may not be enough.

The risks generated from misleading forecasts are typically ignored or downplayed in infrastructure planning, to the detriment of social and economic welfare. Risks, therefore, have a doubly negative effect in this particular type of planning, since it is one thing to take on a risk that one has calculated and is prepared to take, much as insurance companies and professional investors do, while it is quite another matter--that moves risk-taking to a different and more problematic level--to ignore risks. This is especially the case when risks are of the magnitude we have documented here, with many demand forecasts being off by more than 50 percent on investments that measure in hundreds of millions of dollars. Such behavior is bound to produce losers among those financing infrastructure, be they tax payers or private investors. If the losers, or, for future projects, potential losers, want to protect themselves, then our study shows that the risk of faulty forecasts, and related risk assessment and management, must be placed at the core of planning and decision making.

Our goal with this article has been to take a first step in this direction by developing the necessary data and approach.

The policy implications of our findings are clear. First, the findings show that a major planning and policy problem--namely misinformation--exists for this highly expensive field of public policy. Second, the size and perseverance over time of the problem of misinformation indicate that it will not go away by merely pointing out its existence and appealing to the good will of project promoters and planners to make more accurate forecasts. The problem of misinformation is an issue of power and profit and must be dealt with as such, using the mechanisms of transparency and accountability we commonly use in liberal democracies to mitigate rent-seeking behavior and the misuse of power. To the extent that planners partake in rent-seeking behavior and misuse of power, this may be seen as a violation of their code of ethics, that is, malpractice. Such malpractice should be taken seriously by the responsible institutions. Failing to do so amounts to not taking the profession of planning seriously.

## Acknowledgments

The authors wish to thank Don Pickrell, Martin Wachs, the *JAPA* editors, and four anonymous referees for their valuable help. Research for the article was supported by the Danish Transportation Council and Aalborg University, Denmark.

Table 1: Inaccuracy in forecasts of rail passenger and road vehicle traffic.

| | Rail<br><br>[figures in square parentheses<br>include two statistical outliers] | Road |
|---|---|---|
| Average inaccuracy (%) | -51.4 (sd=28.1)<br><br>[-39.5 (sd=52.4)] | 9.5 (sd=44.3) |
| Percentage of projects with inaccuracies larger than ±20% | 84<br><br>[85] | 50 |
| Percentage of projects with inaccuracies larger than ±40% | 72<br><br>[74] | 25 |
| Percentage of projects with inaccuracies larger than ±60% | 40<br><br>[41] | 13 |

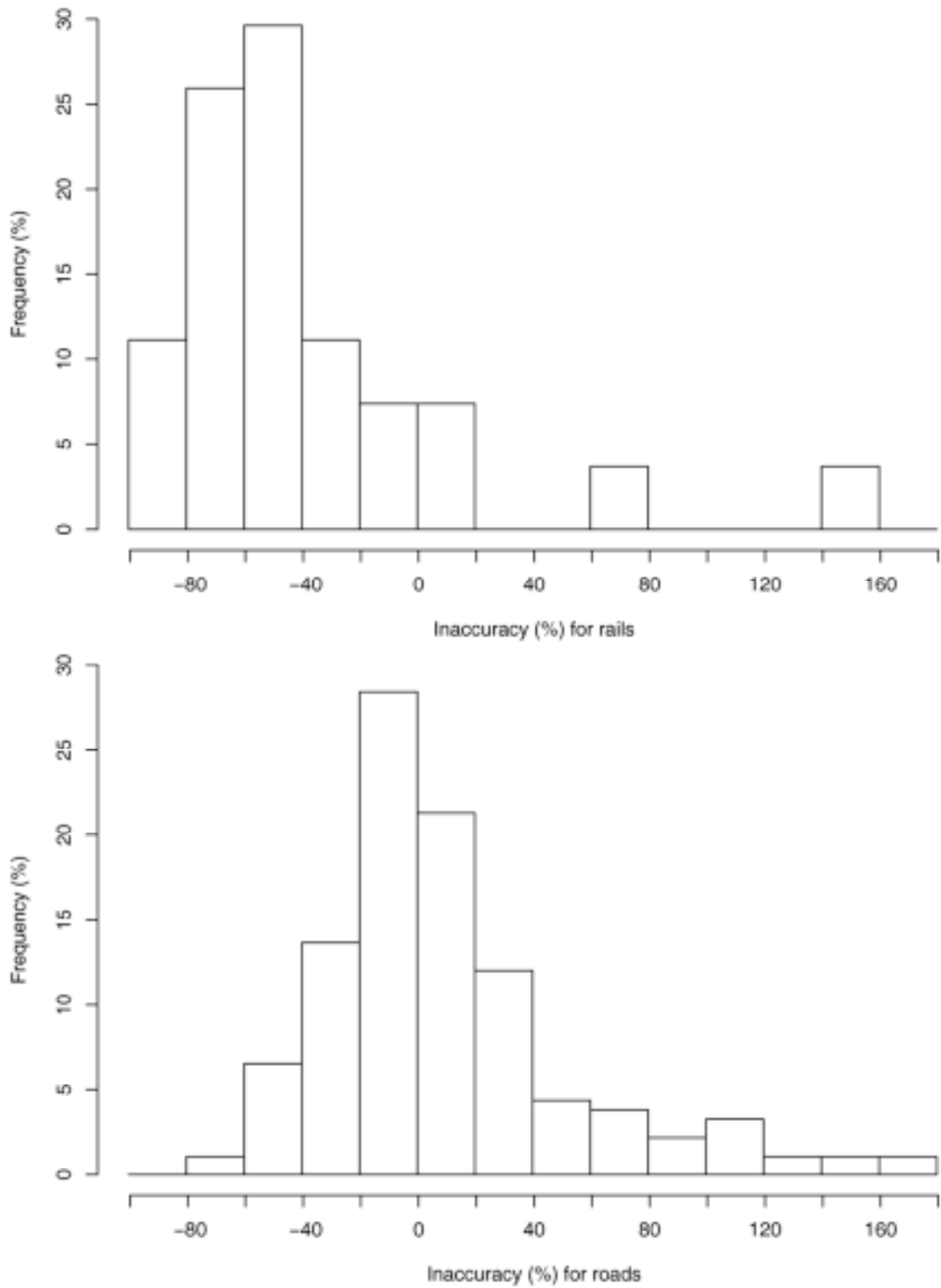

FIGURE 1: Inaccuracies of traffic forecasts in transportation infrastructure projects split into 27 rail and 183 road projects

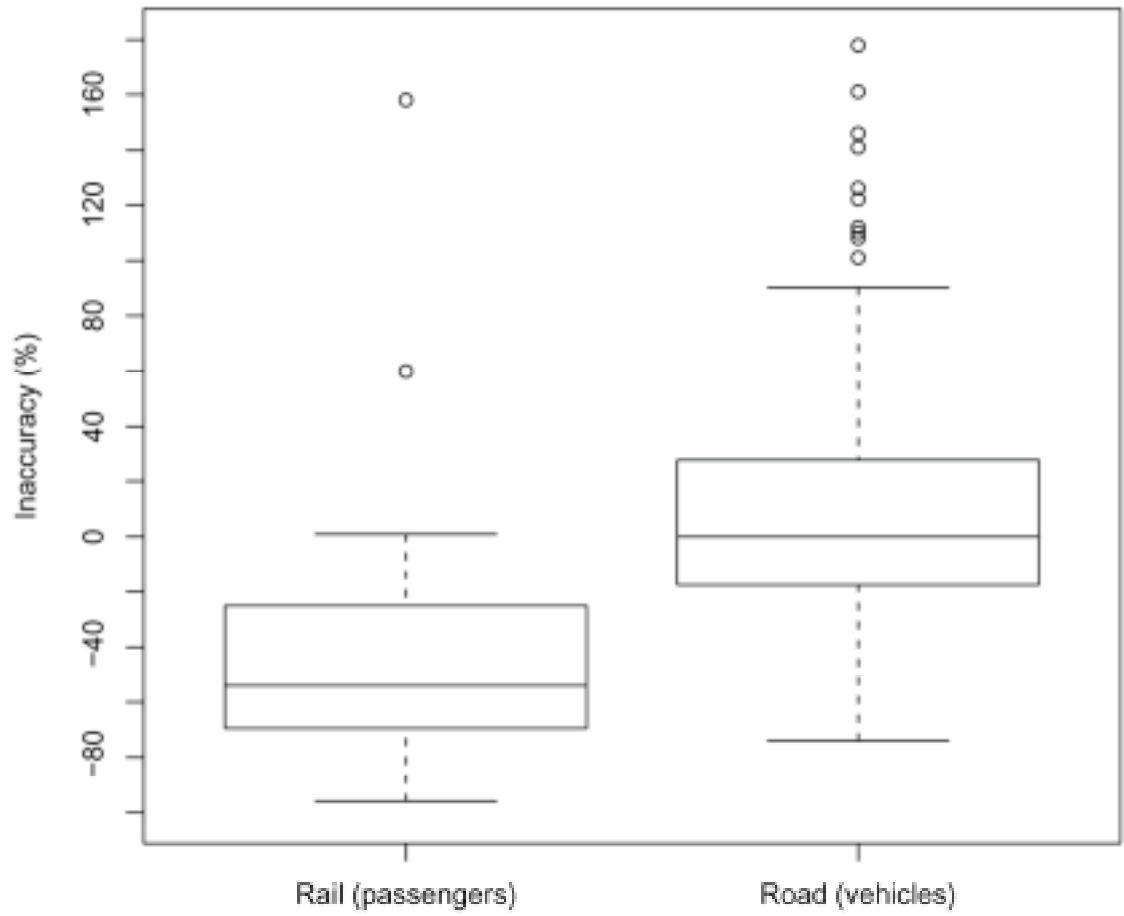

FIGURE 2: Inaccuracies of traffic forecasts in 210 transportation infrastructure projects

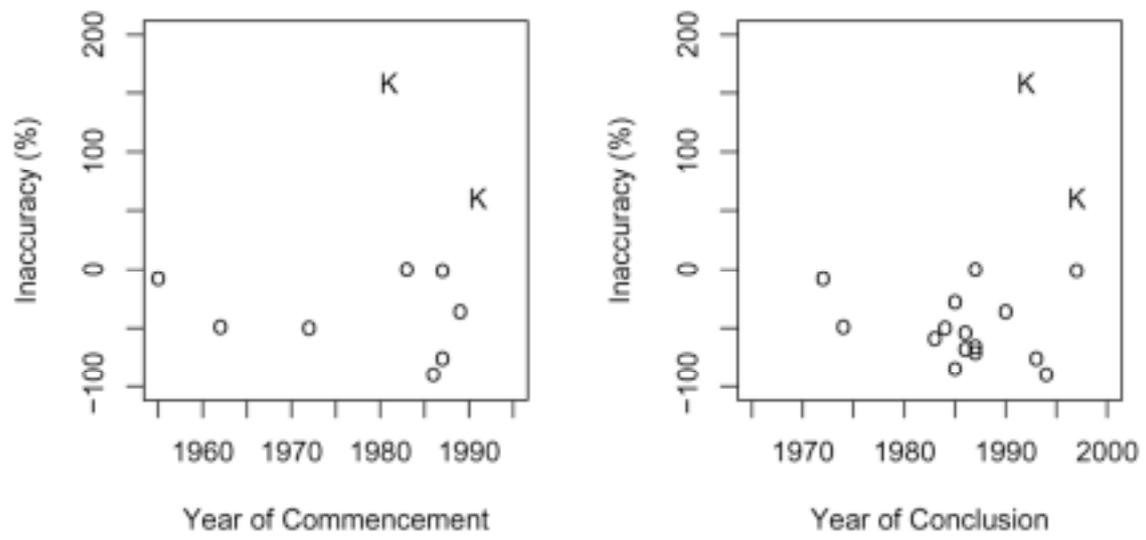

FIGURE 3: Inaccuracy in number of passengers (K = Karlsruhe)

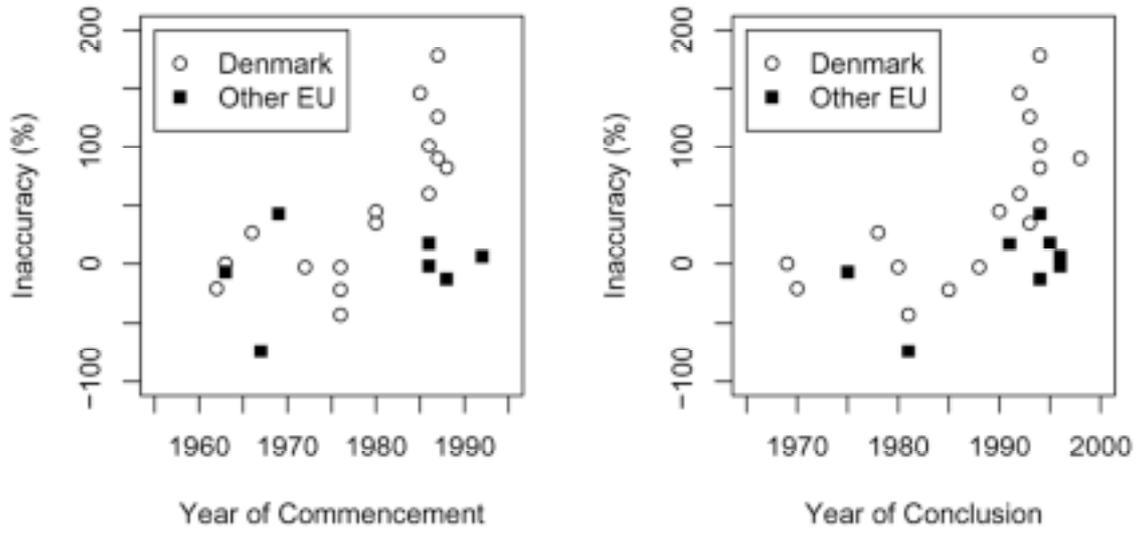

FIGURE 4: Inaccuracy in number of vehicles

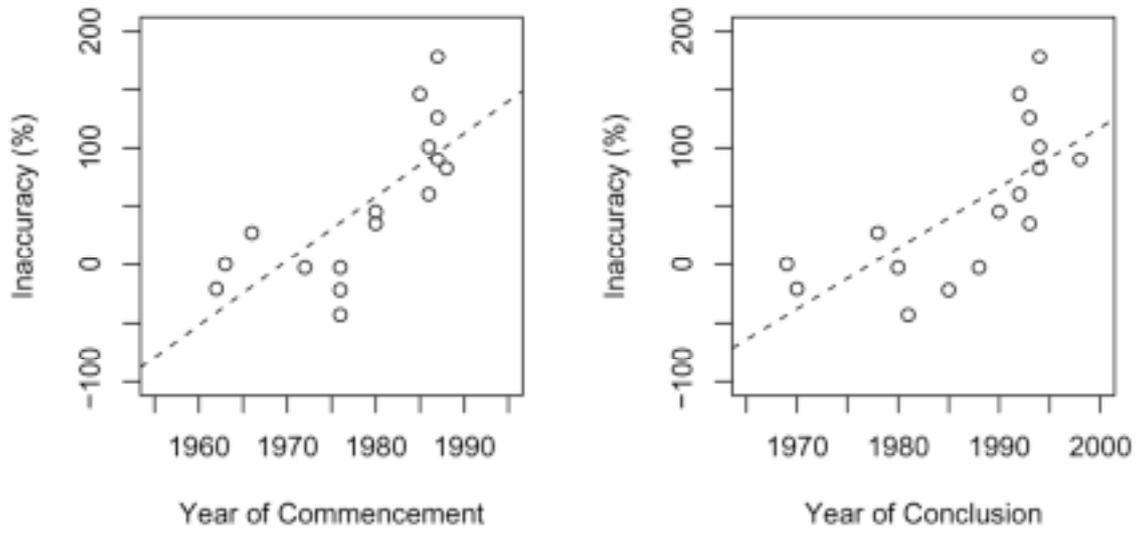

FIGURE 5: Inaccuracy in number of vehicles for Danish projects

Percentage of projects

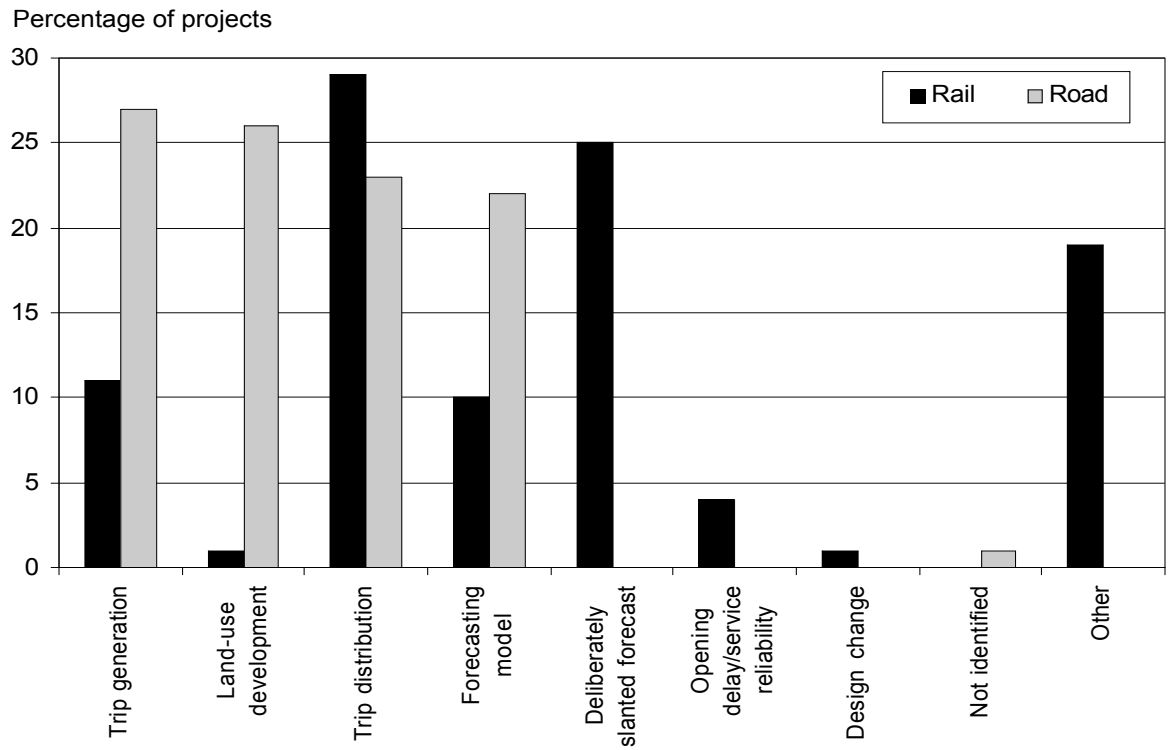

FIGURE 6: Stated causes of inaccuracies in traffic forecasts, 26 rail projects and 208 road projects

# Notes

___________________________

[1] All projects that we know of for which comparable data on forecasted and actual traffic were obtainable were considered for inclusion in the sample. This was 485 projects. 275 projects were then rejected because of unclear or insufficient data quality. More specifically, of the 275 projects rejected, 124 were rejected because inaccuracy had been estimated in ways different from and incomparable to the way we decided to estimate inaccuracy; 151 projects were rejected because inaccuracies for these projects had been estimated on the basis of adjusted data for actual traffic instead of using original, actual count data as we decided to do. All projects for which valid and reliable data were available were included in the sample. This covers both projects for which we ourselves collected the data, and projects for which other researchers in other studies did the data collection. - Our own data collection concentrated on large European projects, because too few data existed for this type of project to allow comparative studies. We collected primary data on the accuracy of traffic forecasts for 31 projects in Denmark, France, Germany, Sweden, and the UK and were thus able to increase many times the number of large European projects with reliable data for both actual and estimated traffic, allowing for the first time comparative studies for this type of project where statistical methods can be applied. Other projects were included in the sample from the following studies: Webber (1976), Hall (1980), National Audit Office (1985), National Audit Office (1988), Fouracre, Allport, and Thomson (1990), Pickrell (1990), Walmsley and Pickett (1992), Skamris (1994), and Vejdirektoratet (1995). Statistical tests showed no differences between data collected through our own surveys and data collected from the studies carried out by other researchers.

[2] The figures mentioned here should be interpreted with caution. Without a published report for the FTA study it is difficult to evaluate the assumptions behind the study and thus the validity and comparability of its results. When the study report has been published, such evaluation should be possible.

[3] We find that the estimated quantities are better than the actual quantities as a measure for project size in the evaluation of inaccuracy, because the estimates are what is known about size at the time of decision to build (and the time of making the forecasts) and using actual quantities would result in the mixing of cause and effect.

[4] As in the other parts of our analyses, here too we include both projects for which we ourselves collected primary data and projects for which other researchers did the data collection as part of other studies, which we then used as secondary sources. Again our own data collection concentrated on large European projects, because data were particularly wanting for this project type. By means of a survey questionnaire and meetings with project managers we collected primary data on causes of inaccurate traffic forecasts for 16

projects, while we collected secondary data for 218 projects from the following studies: Webber (1976), Hall (1980), National Audit Office (1988), Fouracre et al. (1990), Pickrell (1990), Wachs (1990), Leavitt et al. (1993), UK Department of Transportation (1993), Skamris (1994), and Vejdirektoratet (1995).

[5] The closest we have come to an outside view on travel demand forecasts is Gordon and Wilson's (1984) use of regression analysis on an international cross section of light-rail projects to forecast patronage in a number of light-rail schemes in North America.

[6] The lower limit of a one-third share of private risk capital for such capital to effectively influence accountability is based on practical experience. See more in Flyvbjerg, Bruzelius, and Rothengatter (2003, 120-123).